# Electronic structure of the parent compound of superconducting infinite-layer nickelates

M. Hepting[1†], D. Li[1], C. J. Jia[1], H. Lu[1], E. Paris[2], Y. Tseng[2], X. Feng[1], M. Osada[1], E. Been[1], Y. Hikita[1], Y.-D. Chuang[3], Z. Hussain[3], K. J. Zhou[4], A. Nag[4], M. Garcia-Fernandez[4], M. Rossi[1], H. Y. Huang[5], D. J. Huang[5], Z. X. Shen[1], T. Schmitt[2], H. Y. Hwang[1], B. Moritz[1], J. Zaanen[6], T. P. Devereaux[1], and W. S. Lee[1*]

[1]Stanford Institute for Materials and Energy Sciences, SLAC National Accelerator Laboratory and Stanford, Menlo Park, California 94025, USA

[2]Photon Science Division, Swiss Light Source, Paul Scherrer Institut, CH-5232 Villigen PSI, Switzerland

[3]Advanced Light Source, Lawrence Berkeley National Laboratory

[4]Diamond Light Source, Harwell Science and Innovation Campus, Didcot, Oxfordshire OX11 0DE, United Kingdom

[5]NSRRC, Hsinchu Science Park, Hsinchu 30076, Taiwan

[6]Leiden Institute of Physics, Leiden University, 2300 RA Leiden, The Netherland

Correspondence to: leews@stanford.edu

Present address:

†Max Planck Institute for Solid State Research, 70569 Stuttgart, Germany

**The search for oxide materials with physical properties similar to the cuprate high $T_c$ superconductors, but based on alternative transition metals such as nickel, has grown and evolved over time [1-10]. The recent discovery of superconductivity in doped infinite-layer nickelates $R$NiO$_2$ ($R$ = rare-earth element) [11,12] further strengthens these efforts. With a crystal structure similar to the infinite-layer cuprates – transition metal oxide layers separated by a rare-earth spacer layer – formal valence counting suggests that these materials have monovalent Ni$^{1+}$ cations with the same $3d$ electron count as Cu$^{2+}$ in the cuprates. Here, we use x-ray spectroscopy in concert with density functional theory to show that the electronic structure of $R$NiO$_2$ ($R$ = La, Nd), while similar to the cuprates, includes significant distinctions. Unlike cuprates with insulating spacer layers between the CuO$_2$ planes, the rare-earth spacer layer in the infinite-layer nickelate supports a weakly-interacting three-dimensional $5d$ metallic state. This three-dimensional metallic state hybridizes with a quasi-two-dimensional, strongly correlated state with $3d_{x^2-y^2}$ symmetry in the NiO$_2$ layers. Thus, the infinite-layer nickelate can be regarded as a sibling of the rare earth intermetallics [13-15], well-known for heavy Fermion behavior, where the NiO$_2$ correlated layers play an analogous role to the $4f$ states in rare-earth heavy Fermion compounds. This unique Kondo- or Anderson-lattice-like "oxide-intermetallic" replaces the Mott insulator as the reference state from which superconductivity emerges upon doping.**

While the mechanism of superconductivity in the cuprates remains a subject of intense research, early on it was suggested that the conditions required for realizing high $T_c$ superconductivity are rooted in the physics of a two-dimensional electron system subject to strong local repulsion [16, 17]. This describes the Mott (charge-transfer) insulators in the stoichiometric parent compounds, characterized by spin ½ Heisenberg antiferromagnetism, from which superconductivity emerges upon doping. A long-standing question regards whether these "cuprate-Mott" conditions can be realized in other oxides; and extensive efforts to synthesize and engineer nickel oxides (nickelates) have promised such a realization [1-10]. Infinite-layer $NdNiO_2$ became the first such nickelate superconductor following the recent discovery of superconductivity in Sr-doped samples [11]. The undoped parent compound, produced by removing the apical oxygen atoms from the perovskite nickelate $NdNiO_3$ using a metal hydride-based soft chemistry reduction process [10, 18-20], appears to be a close sibling of the cuprates—it is isostructural to the infinite-layer cuprates with monovalent $Ni^{1+}$ cations and possesses the same $3d^9$ electron count as that of $Cu^{2+}$ cations in undoped cuprates. Yet, as we will reveal, the electronic structure of the undoped $R$NiO$_2$ ($R$ = La and Nd) remains distinct from the Mott, or charge-transfer, compounds of undoped cuprates, and even other nickelates.

As a reference, we first discuss the electronic structure of canonical nickelates, NiO and LaNiO$_3$. Rock salt NiO is a charge-transfer insulator, as characterized in the Zaanen-Sawatzky-Allen scheme [21], whose charge-transfer energy $\Delta$ (promoting charge from oxygen ligands to Ni $d$ orbitals) lies below the Coulomb interaction scale $U$ on Ni sites. The valence Ni $d$ orbitals strongly hybridize with oxygen ligands, yielding wavefunctions with mixed character $\alpha|3d^8\rangle + \beta|3d^9L\rangle$ ($\alpha^2 + \beta^2 = 1$), with $\beta^2 \sim 0.2$ [22, 23] per in a NiO$_6$ octahedron, where $L$ denotes a ligand hole on the oxygens. Such Ni-O ligand hybridization gives rise to a pre-peak in x-ray absorption spectroscopy (XAS) near the O $K$-edge (Fig. 1a). In addition, a large band gap set by $\Delta$ appears in the oxygen partial density of states (PDOS) obtained both experimentally (Fig. 1b) and from LDA+U calculations (Fig. 1e). In the perovskite $R$NiO$_3$ where formal valence counting would give $Ni^{3+}$ ($3d^7$), both theoretical and experimental studies indicate that the perovskite structure leads to a decrease of Δ, such that it becomes effectively negative [24]. Under such a scenario, electrons from oxygen ligands spontaneously transfer to Ni cations, giving rise to "self-doped" holes on the ligands, and a pre-peak in the O $K$-edge XAS (Fig. 1a). As expected for a negative charge-transfer metal, no band gap appears in the oxygen PDOS (Figs. 1c and f) [25].

The O $K$-edge XAS tells a very different story for the infinite-layer nickelates LaNiO$_2$ and NdNiO$_2$, as shown in Fig. 1a. The lack of a pre-edge peak suggests that the oxygen ligands carry significantly less weight in the ground state wave function, signaling a weaker effective mixing between oxygen and the Ni$^{1+}$ cations. Unlike NiO and LaNiO$_3$, the oxygen PDOS (Fig. 1d) exhibits a diminished weight near the Fermi energy, also indicating that oxygen $2p$ orbitals carry less weight in the states near the Fermi energy by comparison; all of which is consistent with the calculated oxygen PDOS from LDA+U (Fig. 1g).

While the oxygen electronic structure deviates significantly from other nickelates and cuprates [26], we examine the electronic structure of the Ni cation in $R$NiO$_2$ using both XAS and resonant inelastic x-ray scattering (RIXS) at the Ni $L_3$-edge (a core-level $2p$ to valence $3d$ transition). As shown in Fig. 2a, while XAS for both NiO and LaNiO$_3$ exhibit distinct multi-peak structures originating from $2p^63d^8$–$2p^53d^9$ and $2p^63d^8L^n$–$2p^53d^9L^n$ multiplet transitions, respectively [23, 24], XAS for the infinite-layer nickelates shows a main absorption peak (denoted A), which closely resembles the single peak associated with the $2p^63d^9$–$2p^53d^{10}$ transition in cuprates [27]. In particular for LaNiO$_2$, the XAS exhibits an additional lower energy shoulder A', at which the RIXS spectrum consist a lower energy feature of ~ 0.6 eV (Fig. 2b and e). Note that this feature is absent in the $R$NiO$_3$ ($R$ = La, Nd) compounds (Fig. 2c, e, and Ref. 24). Using exact diagonalization (see Method), we reproduce the general features from XAS and RIXS (Figs. 2f-h), including the A' features, which highlights the hybridization between the Ni

$3d_{x^2-y^2}$ and La $5d$ orbitals. Thus, in configuration interaction, the Ni state can be expressed as a combination of $|3d^9>$ and $|3d^8R>$ where $R$ denotes a charge transfer to the rare-earth cation (See Method and Supplementary Table 2). In NdNiO$_2$, the ~ 0.6 eV feature due to the Nd-Ni hybridization also exists in RIXS (Figs. 2d, e), but it's resonance energy (A') almost coincides with the main absorption peak A. As a consequence, the A' feature cannot be resolved in XAS (Fig. 2a).

To further analyze the electronic structure, we turn to density functional theory. The LDA+U scheme [28] has a long track record of reproducing correctly the gross features of correlated electronic structure for transition metal oxides. While generally first principle, one cannot be certain about the value of the local Coulomb interaction U; however, we can put bounds on it. The infinite-layer nickelates are undoubtedly less good metals than elemental nickel, characterized by U ~ 3 eV, which we can take as a lower bound. From O $K$-edge XAS, the Coulomb interaction should be smaller than that of the large band gap charge-transfer insulator NiO, where U ~ 8 eV. Here, we choose U = 6 eV in our calculations for LaNiO$_2$ (with a lowest energy antiferromagnetic solution, see Method for details), revealing some salient features that correlate with experimental observations: (a) As shown in Fig. 1g (and Fig. 3a), when compared to other nickelates, the oxygen $2p$ bands lie significantly further away from the Fermi energy, signaling reduced oxidation and implying a charge-transfer energy Δ that exceeds *U*. This places the $R$NiO$_2$ infinite-layer nickelates within the Mott-Hubbard regime of the Zaanen-Sawatzky-Allen scheme [21]. (b) The density of states near E$_F$ is dominated by the half-filled Ni $3d_{x^2-y^2}$ states, which appear isolated from the occupied Ni $3d$ bands. The characteristic lower and upper Hubbard bands (Fig. 3a), at least in part, signal a textbook single-band Hubbard model, all but confirming that the Ni cation should be in a very nearly monovalent $3d^9$ state, consistent with the Ni $L$-edge

XAS and RIXS (Fig. 2). (c) The density of states at $E_F$ is actually finite, but small, as shown upon closer inspection of both Figs. 3a and 3b. Near the Γ point, a Fermi surface pocket forms of mainly La $5d$ character (Fig. 3b); it is quite extended and three-dimensional (see the wavefunction at $k_F$, Fig. 3c, and Fermi surface, Fig. 3d). This contrasts with the two-dimensional (2D) nature of the correlated $3d_{x^2-y^2}$ Ni states (Fig. 3b). In other words, the electronic structure of the infinite layer nickelate consists of a low density three-dimensional (3D) metallic rare-earth band coupled to a 2D Mott system.

A minimal model for these materials would look like

$$H = \sum_{k,\sigma}\left(\varepsilon_k^R n_{k,\sigma}^R + \varepsilon_k^{Ni} n_{k,\sigma}^{Ni}\right) + U\sum_i n_{i,\uparrow}^{Ni} n_{i,\downarrow}^{Ni} + \sum_{k,i,\sigma}\left(V_{k,i} c_{k,\sigma}^+ d_{i,\sigma} + h.c.\right),$$

where the first term describes the non-interacting rare-earth ($R$) and Ni bands with energies $\varepsilon_k^R$ and $\varepsilon_k^{Ni}$, respectively, the second term represents the usual on-site Hubbard interaction with strength $U$ in the quasi-two-dimensional Ni layer, and the third term describes the coupling with strength $V_{k,i}$ between the $R$ and Ni subsystems. Here, $n_{k,\sigma}^R$ and $n_{k,\sigma}^{Ni}$ represent the usual number operators for the $R$ and Ni subsystems, while $c_{k,\sigma}^\dagger$ ($c_{k,\sigma}$) and $d_{k,\sigma}^\dagger$ ($d_{k,\sigma}$) create (annihilate) electrons in the 3D metallic $R$ and 2D Hubbard-like Ni subsystems, respectively. This model resembles the Anderson-lattice (or Kondo-lattice) model for the rare-earth intermetallics [13-15], but with the notable addition of a weakly hybridized single-band Hubbard-like model for the Ni layer, rather than strongly interacting $4f$ states (or localized spin moments). We can take this a step further by "downfolding" the band structure to a minimal model that should be the starting point for the unusual correlated electron physics in this system. Figure 4a shows the band structure for LaNiO$_2$ obtained from LDA without a Hubbard $U$ (see Supplementary Information for details). Here, consistent with previous calculations [2], two bands cross the Fermi level: a fully three-dimensional band with predominantly La $5d$ character, and a quasi-two-dimensional band with Ni

$3d$ character. Wannier downfolding [29] produces one extended orbital with $d_{3z^2-r^2}$ symmetry centered on La (Fig. 4b) and another orbital confined primarily to the NiO$_2$ planes with $d_{x^2-y^2}$ symmetry centered on Ni (Fig. 4c), which are fully consistent with the expected orbital arrangements given the crystal and ligand field symmetries for this material and the LDA+U results shown in Fig. 3. Full details about the downfolded model, including effective model parameters, can be found in the Supplementary Table 3.

This downfolded model is to the best of our knowledge unique to this particular system. Viewed theoretically, this is uncharted territory and it is a natural question to ask what happens to the basic single-band Hubbard model when its states weakly hybridize with a metallic band. For example, do the spins in the NiO$_2$ layers order antiferromagnetically or will the Kondo effect strongly screen the local moments and give rise to electronic band hybridizations in analogy to the case of heavy Fermions [13-15]? Note that unlike the rare-earth intermetallics, here, Ni spins interact via the strong short range super-exchange interaction, which replaces the RKKY interactions in the heavy Fermion compounds. More importantly, can superconductivity emerge in this model by introducing doped charge carriers? Apparently, experimental information, particularly about the Fermi surface, magnetic susceptibility, and information about other elementary excitations, such as spin, charge, and phonon excitations, will be required to gain further insights. Nevertheless, our results have provided a first glimpse into the novel electronic structure of the parent compounds of superconducting infinite-layer nickelates, which appear to serve as a birthplace of superconductivity upon doping.

## Methods

**Materials**

$LaNiO_3$ films with 12 and 50 nm thicknesses were grown on top of 5 × 5 mm$^2$ $TiO_2$-terminated $SrTiO_3$ (001) substrates by pulsed laser deposition (PLD) using a 248nm KrF excimer laser. Prior to growth, $SrTiO_3$ substrates were pre-annealed at an oxygen partial pressure ($p_{O2}$) of 5 × 10$^{-6}$ torr for 30 min at 950 °C to achieve sharp step-and-terrace surfaces. The films were subsequently grown at a substrate temperature $T_g$ of 575 °C and $p_{O2}$ = 34 mtorr, using 1.4 J cm$^{-2}$ laser fluence and 4 mm$^2$ laser spot size on the target. The growth was monitored by reflection high-energy electron diffraction (RHEED) intensity oscillations. After the growth, the samples were cooled down to room temperature in the same oxygen environment. Characterization by x-ray diffraction (XRD) scans with Cu Kα radiation indicated the presence of the perovskite phase of (001)-oriented $LaNiO_3$ and high epitaxial quality for all as-grown films. AFM topographic scans showed atomically flat surfaces. Reducing conditions [30] were adapted to remove apical oxygen for producing both the (001)-oriented $LaNiO_{2.5}$ and $LaNiO_2$ phases. For reduction experiments, each $LaNiO_3$ sample was cut into 2 pieces of 2.5 × 5 mm$^2$ size. The 2.5 × 5 mm$^2$ sample was then vacuum-sealed together with blocks of $CaH_2$ powder in a Pyrex glass tube (pressure < 0.1 mtorr). The tube was heated to 240 °C at a rate of 10 °C/min and kept at this temperature for 30-120 mins, before cooled down to room temperature at a rate of 10 °C/min. After the annealing process, remnant $CaH_2$ powder on sample surface was rinsed off by 2-Butanone. The XRD scans in Supplementary Fig. 1a show the characteristic Bragg peaks of the 12 nm $LaNiO_3$ film and the ~ 50 nm $LaNiO_2$ film used in the XAS and RIXS measurements of the main text. Additionally, a ~ 50 nm $LaNiO_{2.5}$ film was characterized as a reference sample. The 2θ peak position of these three films coincide with that of similar films on SrTiO$_3$ [30]. The *c*-axis lattice constants extracted from the XRD scans are 3.809, 3.771, and 3.407 Å for the LaNiO$_2$, LaNiO$_{2.5}$, and LaNiO$_2$ film, respectively. In comparison to LaNiO$_2$ powder [17, 30], the *c*-axis lattice constant of the film is slightly expanded due to the compressive strain induced by the SrTiO$_3$ substrate.

NdNiO$_2$ films grown on a SrTiO$_3$ substrate with a thickness of ~10 nm were prepared using the conditions described in Ref. 10. NdNiO$_2$ films with and without a capping layer of 5 unit cell (u.c.) SrTiO$_3$ were measured, which show the same spectral properties. As a reference, we also measured a NdNiO$_3$ film grown on a SrTiO$_3$ substrate and capped with a 5 u.c. SrTiO$_3$ film.

Supplementary Fig. 1b displays the resistivity as a function of temperature of the LaNiO$_3$ film in a four-probe geometry, which shows metallic behaviour down to 2 K. The LaNiO$_2$ film exhibits higher resistivity than LaNiO$_3$ at 300 K, which increases further with decreasing temperature [Supplementary Fig. 1b]. Similar transport properties were reported in Refs. 19, 30, 31.

Commercially available NiO powder with ≥ 99.995% purity (Sigma-Aldrich) was used for the measurements.

**XAS and RIXS measurements**

XAS and RIXS spectra of the La-based nickelate samples were measured at the ADRESS beamline with the SAXES spectrometer at the Swiss Light Source (SLS) of the Paul Scherrer Institute [32]. For the RIXS measurements the scattering angle was fixed to 130° and the combined instrument resolution was approximately 100 meV at the Ni $L_3$-edge. The scattering plane coincided with the crystallographic *ac* (*bc*) plane with a grazing incident angle θ = 15°. The XES and RIXS spectra shown in Fig. 1 of the main text were measured with π-polarized incident photons. Due to the strong fluorescence signal from the STO substrate, the XES of LaNiO$_3$ and LaNiO$_2$ shown in Fig.

1 were obtained from the fluorescence signal identified in RIXS incident-photon-energy-and emission-energy map across the oxygen pre-edge (incident photon energy from ~ 525 eV to ~ 530 eV). The elastic line and weak Raman-like excitations (in LaNiO$_2$) were removed for clarity.

The XAS and XES spectra of NiO shown in Fig. 1a were measured at beamline BL8.0 using the q-RIXS endstation of the Advanced Light Source (ALS) of the Lawrence Berkeley National Laboratory. For the RIXS/XES measurements the scattering angle was fixed to 130° and the combined instrument resolution was approximately 300 meV at the Ni $L_3$-edge and approximately 200 meV at the O $K$-edge. The XAS at the O $K$-edge for NdNiO$_3$ and NdNiO$_2$ (Fig. 1a) were taken at 41A BlueMagpie beamline at Taiwan Photon Source. The XAS and RXIS map at the Ni $L$-edge for the NdNiO$_2$ were taken at I21 beamline at the Diamond Light Source. The RIXS spectrometer is set at 146 degree, with a resolution of approximately 50 meV. The scattering plane coincided with the crystallographic *ac* (*bc*) plane with a grazing incident angle ~10°. π-polarized incident photons were used for this measurement.

All XAS at O $K$-edge (Fig. 1) were taken in fluorescence yield mode with a grazing incident angle of 10 and 20 degrees for the La-based and Nd-based nickelates, respectively. The grazing incident geometry is used to reduce the signal arising from the STO substrate. The spectrum is normalized such that the intensity at the pre-edge and the post-edge is 0 and 1, respectively.

All XAS at the Ni $L$-edge (Fig. 2) were taken in fluorescence yield mode with a normal incident geometry. These XAS are normalized such that the intensity at the pre-edge and the post-edge are 0 and 1, respectively. For the XAS of the La-based nickelates, the intense La $M_4$-line centered around 850.5 eV (Supplementary Fig. 2a) was fitted by a Lorentzian peak profile and subtracted from the LaNiO$_3$ and LaNiO$_2$ XAS to correct for the overlap between the tail of the La $M_4$-line and the Ni $L_3$-edge. The resulting spectra are shown in Fig. 2 of the main text.

**Theory calculations**

For the oxygen partial density of states (PDOS) as shown in Fig. 1**e**-**g** and the electronic structure of LaNiO$_2$ shown in Figure 3, LDA+U calculations were performed using the GGA method and the simplified version by Cococcioni and de Gironcoli [33], as implemented in Quantum ESPRESSO [34]. We find that an antiferromagnetic solution, with wave vector ($\pi,\pi,\pi$), leads to the lowest energy, with a two Ni, body centered tetragonal (BCT) unit cell and corresponding Brillouin zone.

The Ni $L_3$-edge RIXS calculations [Fig. 2] were performed using an exact diagonalization technique [35, 36], which accounts for the full overlap of the many-body wavefunctions. The microscopic Hamiltonian used for these calculations includes both material-specific on-site energies and hybridizations as encoded in a Wannier downfolding of the bandstructure [37] and the full set of Coulomb interactions as expressed in terms of Slater integrals. The Wannier downfolding parameters for paramagnetic LaNiO$_2$ (a one Ni, tetragonal unit cell), as shown in Supplementary Table 1 were obtained from Wannier90 [38] for 12-orbital (O $p_x/p_y/p_z$, Ni $d_{z^2}/d_{x^2-y^2}/d_{xy}/d_{xz}/d_{yz}$, La $d_{z^2}$) Wannier downfolding was used in the Ni $L$-edge RIXS calculation for LaNiO$_2$, where the relevant parameters appear in Supplementary Table 1. The Slater integrals for Ni $3d$ in the LaNiO$_2$ calculations were: $F^0 = 0.5719$ eV, $F^2 = 11.142$ eV, and $F^4 = 6.874$ eV. The Slater integrals for Ni $3d$-$2p$ interactions are: $F^0_{p,d} = 0.148$ eV, $F^2_{p,d} = 6.667$ eV, $G^1_{p,d} = 4.922$ eV, and $G^3_{p,d} = 2.796$ eV. The values of $F^2$, $F^4$, $F^2_{p,d}$, $G^1_{p,d}$ and $G^3_{p,d}$ are taken from Ref [37]. We take 0.7 as a screening factor for the non-monopole terms. A core-level spin-orbit coupling of 12.5 eV has been used for the Ni $2p$ core electrons. The resulting weight of the Ni wave function is shown in Supplementary Table 2.

The two-orbital, low energy model for the physics of LaNiO$_2$ appears in Fig. 4. This model, obtained once again by Wannier downfolding the DFT paramagnetic solution for LaNiO$_2$ in the one Ni, tetragonal unit cell (the same method as that used to obtain the noninteracting part of the Hamiltonian for the LaNiO$_2$ RIXS calculation, but only for the two bands that cross $E_F$), yields the independent hopping parameters listed in Supplementary Table 3, cutoff for absolute values smaller than 0.008 eV. The two Wannier orbitals are shown in Fig. 4b of the main text: (1) a very extended orbital, centered on La, with $d_{3z^2-r^2}$ character, which makes-up the majority character of the three-dimensional band; and (2) a more localized orbital, centered on Ni and primarily confined to the NiO$_2$ plane, with $d_{x^2-y^2}$ character, which makes-up the majority character of the quasi-two-dimensional band. Note that this paramagnetic solution in the tetragonal Brillouin zone has one large quasi-two-dimensional hole-like Fermi surface from the Ni-centered orbital and two smaller three-dimensional electron-like Fermi surfaces center at the Γ- and A-points from the La-centered orbital. The low energy, antiferromagnetic bandstructure from LDA+U [Fig. 3] would result from a (π,π,π) band-folding of the La-centered band, which moves the A-point to the Γ-point, formation of upper and lower Hubbard Ni-centered bands, gapping-out the large hole Fermi surface, and a shift in chemical potential to compensate for the loss of carriers, which leaves a single electron pocket at the Γ-point.

The non-interacting bands of the effective low energy model can be written in tight-binding form as

$$\begin{aligned}
\varepsilon_k^R = \varepsilon_0^R &+ 2\, t_{[0,0,1]}^R \cos(k_z) + 2\, t_{[0,0,2]}^R \cos(2 k_z) + 2\, t_{[0,0,3]}^R \cos(3 k_z) \\
&+ 2\, t_{[1,0,0]}^R [\cos(k_x) + \cos(k_y)] \\
&+ 4\, t_{[1,0,1]}^R [\cos(k_x) + \cos(k_y)] \cos(k_z) \\
&+ 4\, t_{[1,0,2]}^R [\cos(k_x) + \cos(k_y)] \cos(2 k_z) \\
&+ 4\, t_{[1,1,0]}^R \cos(k_x) \cos(k_y) \\
&+ 8\, t_{[1,1,1]}^R \cos(k_x) \cos(k_y) \cos(k_z) \\
&+ 8\, t_{[1,1,2]}^R \cos(k_x) \cos(k_y) \cos(2 k_z) \\
&+ 8\, t_{[1,1,3]}^R \cos(k_x) \cos(k_y) \cos(3 k_z) \\
&+ 4\, t_{[2,0,1]}^R [\cos(2 k_x) + \cos(2 k_y)] \cos(k_z) \\
&+ 8\, t_{[2,1,1]}^R [\cos(2 k_x) \cos(k_y) + \cos(k_x) \cos(2 k_y)] \cos(k_z)
\end{aligned}$$

$$\begin{aligned}
\varepsilon_k^{Ni} = \varepsilon_0^{Ni} &+ 2\, t_{[1,0,0]}^{Ni} [\cos(k_x) + \cos(k_y)] \\
&+ 4\, t_{[1,1,0]}^{Ni} \cos(k_x) \cos(k_y) \\
&+ 2\, t_{[2,0,0]}^{Ni} [\cos(2 k_x) + \cos(2 k_y)] \\
&+ 2\, t_{[0,0,1]}^{Ni} \cos(k_z) \\
&+ 8\, t_{[1,1,1]}^{Ni} \cos(k_x) \cos(k_y) \cos(k_z)
\end{aligned}$$

where the appropriate matrix elements can be found in Supplementary Table 3.

**Acknowledgments**

This work is supported by the U.S. Department of Energy (DOE), Office of Science, Basic Energy Sciences, Materials Sciences and Engineering Division, under contract DE-AC02-76SF00515. X. F. and D. L. acknowledge partial support from the Gordon and Betty Moore Foundation's EPiQS Initiative through Grant GBMF4415. Part of the synchrotron experiments have been performed at the ADRESS beamline of the Swiss Light Source (SLS) at the Paul Scherrer Institut (PSI). The work at PSI is supported by the Swiss National Science Foundation through the NCCR MARVEL (Research Grant 51NF40_141828) and the Sinergia network Mott Physics Beyond the Heisenberg Model - MPBH (Research Grant CRSII2_160765/1). Part of research was conducted at the Advanced Light Source, which is a DOE Office of Science User Facility under contract DE-AC02-05CH11231. We acknowledge preliminary XAS characterization at BL13-3, SSRL by J. S. Lee in the early stage of the project.

## Author contributions

W.S.L., M.H. and H.Y.H. conceived the experiment. M.H., H.L., E.P., Y.T., T.S., and W.S.L. conducted the experiment at SLS. H.L., W.S.L., Z.H. and Y.D.C. conducted the experiment at ALS, USA. H.L, W.S.L, A.N. and K.J.Z. conducted XAS measurement at Diamond Light Source, UK. M.R., W.S.L., H.H., and D.J.H. conducted XAS measurements at NSRRC, Taiwan. J.S.L. contributed to XAS characterization of samples at an early stage of the work. M.H., H.L. and W.S.L. analyzed the data. C.J.J., B.M., J. Z. and T.P.D. performed the theoretical calculations. D.L., X.F., Y.H., and H.Y.H. synthesized and characterized the nickelate samples using transport and XRD. M.H., Z.X.S. and W.S.L. prepared and aligned samples for x-ray spectroscopy measurements. M.H., B.M. J. Z. and W.S.L. wrote the manuscript with input from all authors.

## Competing financial interests

The authors declare no competing financial interests.

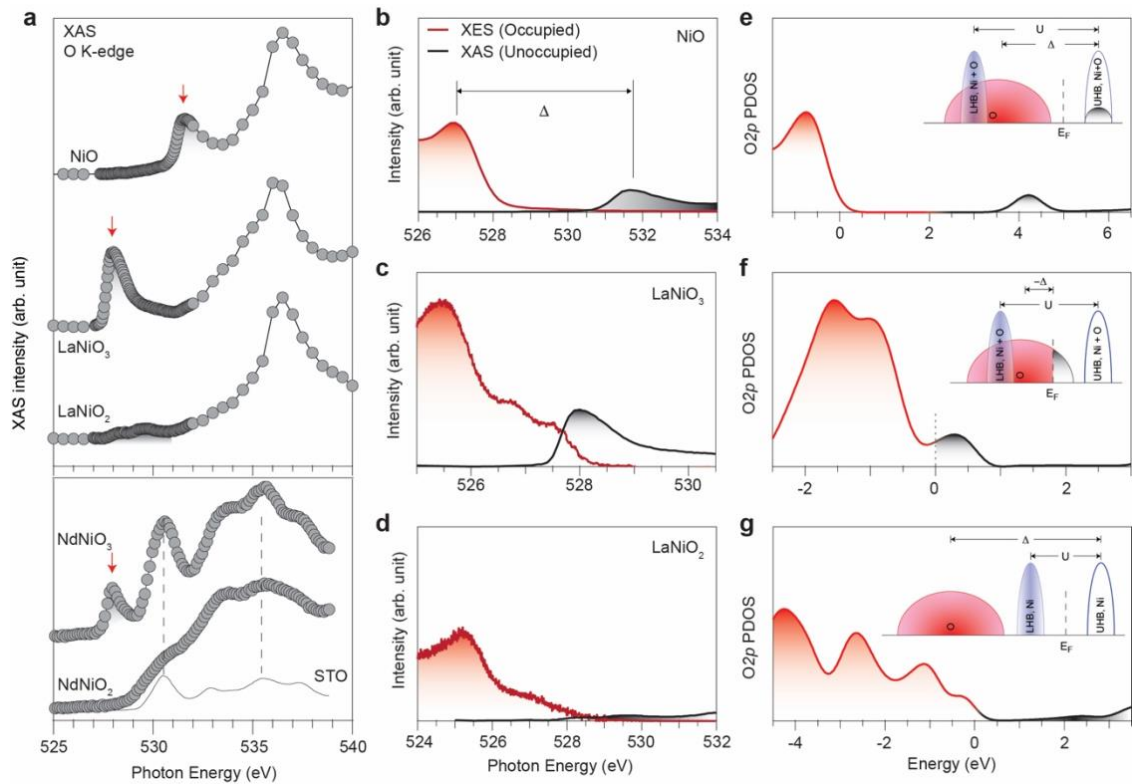

**Figure 1 | X-ray spectroscopy near the O *K*-edge and LDA+U calculation. a**, X-ray absorptions spectra (XAS) of NiO, LaNiO$_3$, and LaNiO$_2$. Red arrows mark the pre-edge peaks indicative of Ni-O hybridization. The lower panel shows the XAS of NdNiO$_3$ and NdNiO$_2$. Dashed vertical lines indicate features of the SrTiO$_3$ (STO) substrate (solid grey line) in the XAS of NdNiO$_3$ and NdNiO$_2$, due to thinner film thickness than that of the La-based films shown in the upper panel. Spectra are vertically offset for clarity. **b-d**, X-ray emission spectrum (XES) and XAS in the pre-edge region, roughly reflecting the occupied (red shading) and unoccupied (black shading) oxygen PDOS, respectively. Vertical lines illustrate the band gap projected in the oxygen density of states, corresponding to the effective charge transfer energy Δ in NiO and LaNiO$_3$. **e-g**: LDA+U calculations for the PDOS with O *2p* orbital character. (U=8eV for NiO and LaNiO$_3$, U=6eV for LaNiO$_2$). Red and black shadings indicate the occupied and unoccupied oxygen PDOS, respectively. Insets are sketches of the relationship between U and Δ in the Zaanen-Sawastzky-Allen scheme for each compound.

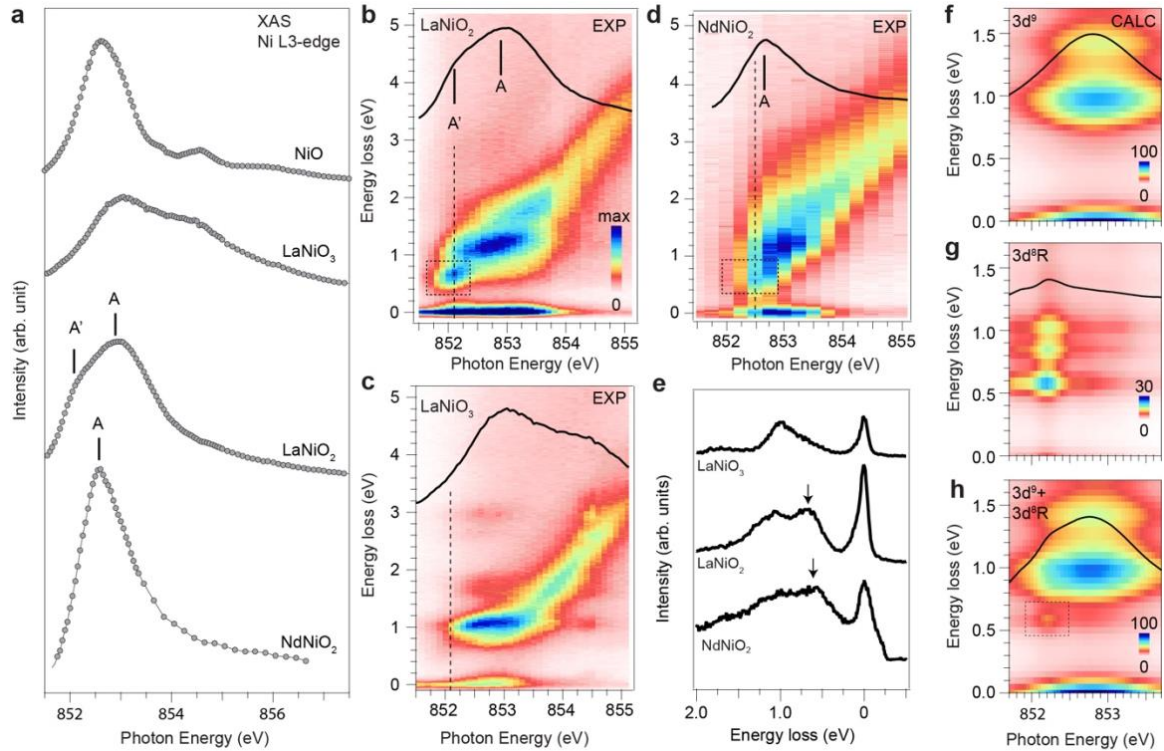

**Figure 2 | X-ray absorption spectroscopy (XAS) and resonant inelastic x-ray scattering (RIXS) at the Ni $L_3$-edge. a**, Normalized absorption spectra across the Ni $L_3$-edge of NiO, $LaNiO_3$, $LaNiO_2$, and $NdNiO_2$. The La $M_4$-line was subtracted from the $LaNiO_2$ and $LaNiO_3$ spectra (see Supplementary Fig. 2). The markers A indicate the main peak for $LaNiO_2$ and $NdNiO_2$. A' labels a lower energy shoulder in the XAS of $LaNiO_2$. Spectra are vertically offset for clarity. **b-d,** RIXS intensity map of $LaNiO_2$, $LaNiO_3$, and $NdNiO_2$ measured as a function of incident photon energy at $T = 20$ K. The corresponding XAS is superimposed as a solid black line in each map. The dashed box highlights the ~0.6 eV features of $LaNiO_2$ and $NdNiO_3$ that are associated with the Ni-La and Ni-Nd hybridization, respectively. **e**, RIXS energy loss spectra of $LaNiO_3$, $LaNiO_2$, and $NdNiO_2$ at incident energies indicated by vertical dashed lines in **b-d**. The black arrows highlight the 0.6 eV features of $LaNiO_2$ and $NdNiO_2$. **f-g,** Calculated RIXS maps and absorption spectra (solid black lines) of $LaNiO_2$ for a $3d^9$, $3d^8R$, and $3d^9+3d^8R$ (R denote a charge transfer to the rare-earth cation) ground state, respectively. The lightly dashed box in panel **f** highlights the same feature as the box in panel **b**.

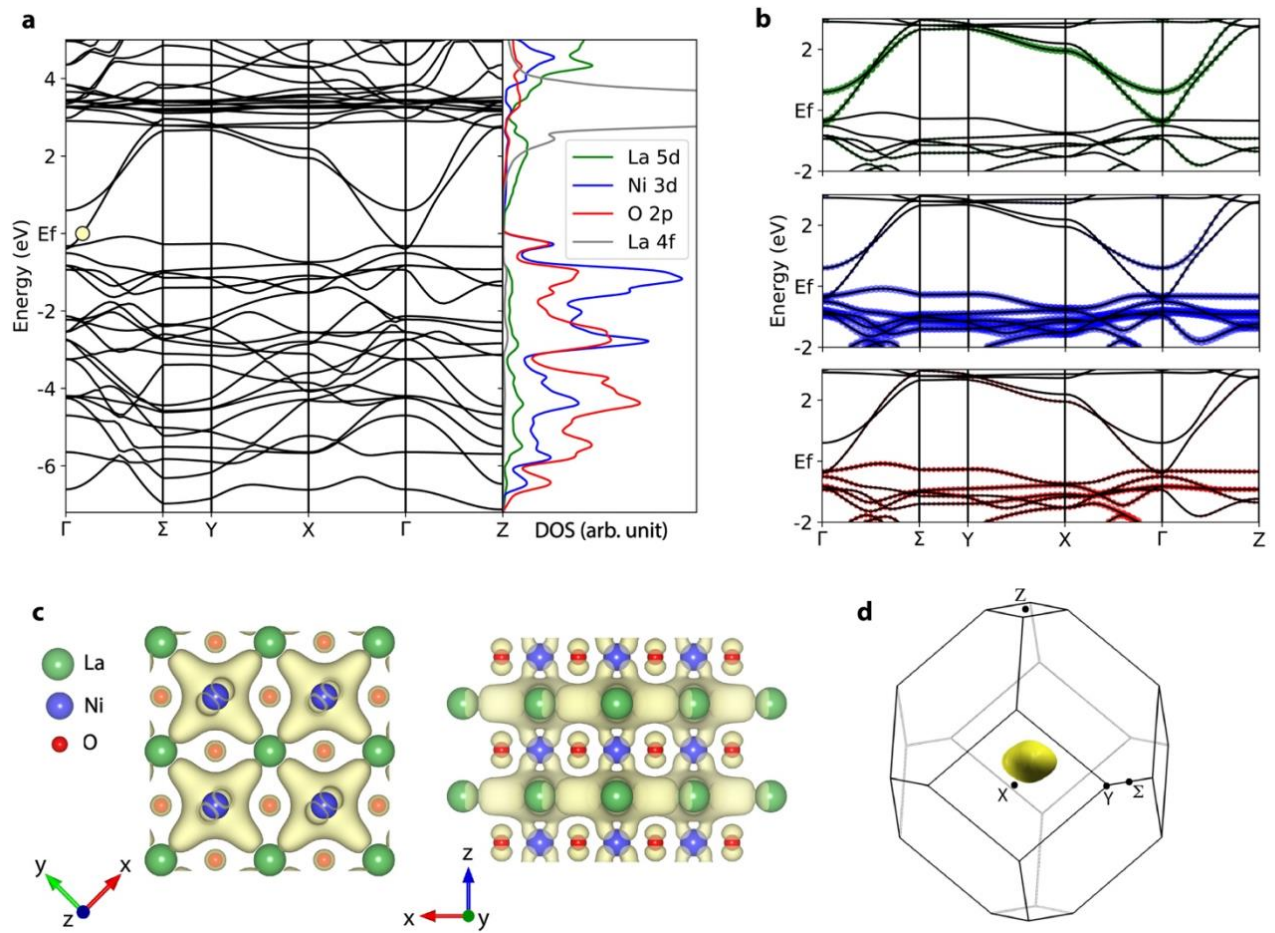

**Figure 3 | Electronic structure of LaNiO₂.** Theoretical calculations of the electronic structure in the LDA+U framework with $U = 6$ eV (antiferromagnetic solution). **a,** Band structure of LaNiO$_2$ along high symmetry directions in the body centered tetragonal (BCT) Brillouin zone (BZ). The BZ with labeled high symmetry points is also shown in **d**. The right-hand side shows the La 5$d$ (green), Ni 3$d$ (blue), O 2$p$ (red) and La 4$f$ (grey) partial density of states with a smaller energy broadening than that used in Figs. 1e-g. **b,** Orbital-projected band structure of LaNiO$_2$ near $E_F$. The color code is identical to that used in panel a, representing the projection onto orbitals with different atomic character. **c**, Top- and side-views of an electron density contour for the single-particle wavefunction at $k_F$ along Γ-Σ (yellow marker in panel **a**). **d**, Fermi surface (closed electron pocket) around Γ with dominant La 5$d$ character in the first BCT BZ with labeled high symmetry points.

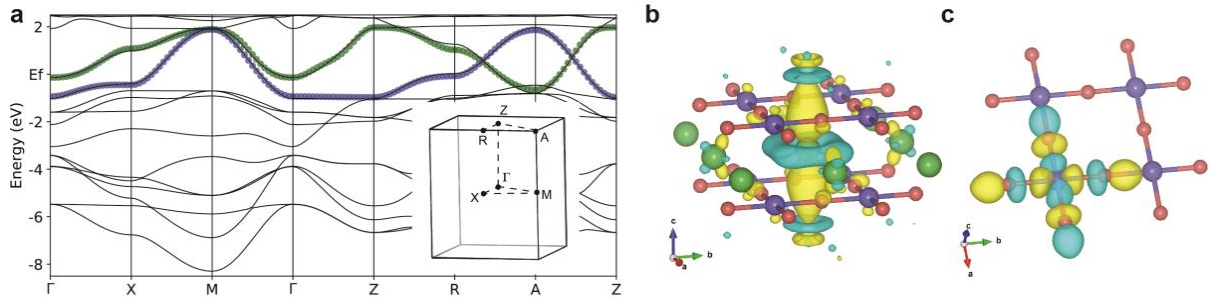

**Figure 4 | Modeling the rare-earth infinite layer nickelates. a**, Band dispersion of $LaNiO_2$, highlighting two bands which cross $E_F$ in the paramagnetic LDA calculation. The inset shows the high symmetry points in the tetragonal BZ. **b**, Isosurface plots for an extended La-centered $d_{3z^2-r^2}$-like and essentially planar Ni-centered $d_{x^2-y^2}$-like Wannier orbital for the minimal low-energy model of $LaNiO_2$. These two orbitals produce the three-dimensional band (La, green) and quasi-two-dimensional band (Ni, blue) highlighted in **a**.

# Supplementary Information for Electronic structure in the parent compound of superconducting infinite layer nickelates

M. Hepting[1,7], D. Li[1], C. J. Jia[1], H. Lu[1], E. Paris[2], Y. Tseng[2], X. Feng[1], M. Osada[1], E. Been[1], Y. Hikita[1], Y.-D. Chuang[3], Z. Hussain[3], K. J. Zhou[4], A. Nag[4], M. Garcia-Fernandez[4], M. Rossi[1], H. Y. Huang[5], D. J. Huang[5], Z. X. Shen[1], T. Schmitt[2], H. Y. Hwang[1], B. Moritz[1], J. Zaanen[6], T. P. Devereaux[1], and W. S. Lee[1*]

[1]Stanford Institute for Materials and Energy Sciences, SLAC National Accelerator Laboratory and Stanford, Menlo Park, California 94025, USA

[2]Photon Science Division, Swiss Light Source, Paul Scherrer Institut, CH-5232 Villigen PSI, Switzerland

[3]Advanced Light Source, Lawrence Berkeley National Laboratory

[4]Diamond Light Source, Harwell Science and Innovation Campus, Didcot, Oxfordshire OX11 0DE, United Kingdom

[5]NSRRC, Hsinchu Science Park, Hsinchu 30076, Taiwan

[6]Leiden Institute of Physics, Leiden University, 2300 RA Leiden, The Netherland

Current affiliation:

[7]Max-Planck-Institute for Solid State Research, Heisenbergstraße 1, 70569 Germany

Correspondence to: leews@stanford.edu

This Supplementary Information Contains:

Supplementary Fig. 1-2

Supplementary Table 1-3

Supplementray Reference 30 - 38

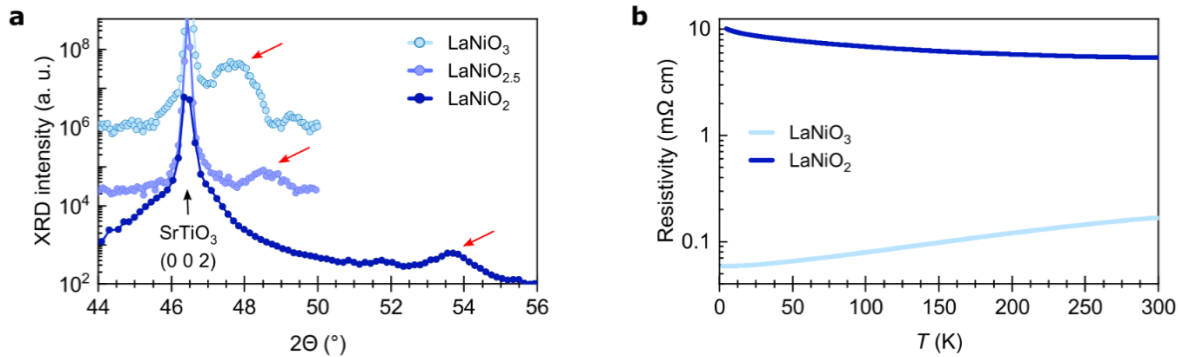

**Supplementary Figure 1 | XRD characterization and electrical transport measurements. a,** XRD pattern of the LaNiO$_3$, LaNiO$_{2.5}$, and LaNiO$_2$ films grown on SrTiO$_3$ (001) substrates, measured with Cu K$\alpha$ radiation. The red arrows indicate the nickelate film peaks and the black arrows the (002) SrTiO$_3$ substrate peak. The film peak shifts to higher 2$\theta$ values as a function of apical oxygen reduction. The curves are offset in vertical direction for clarity. **b,** Resistivity vs. temperature of the LaNiO$_3$ and LaNiO$_2$ film.

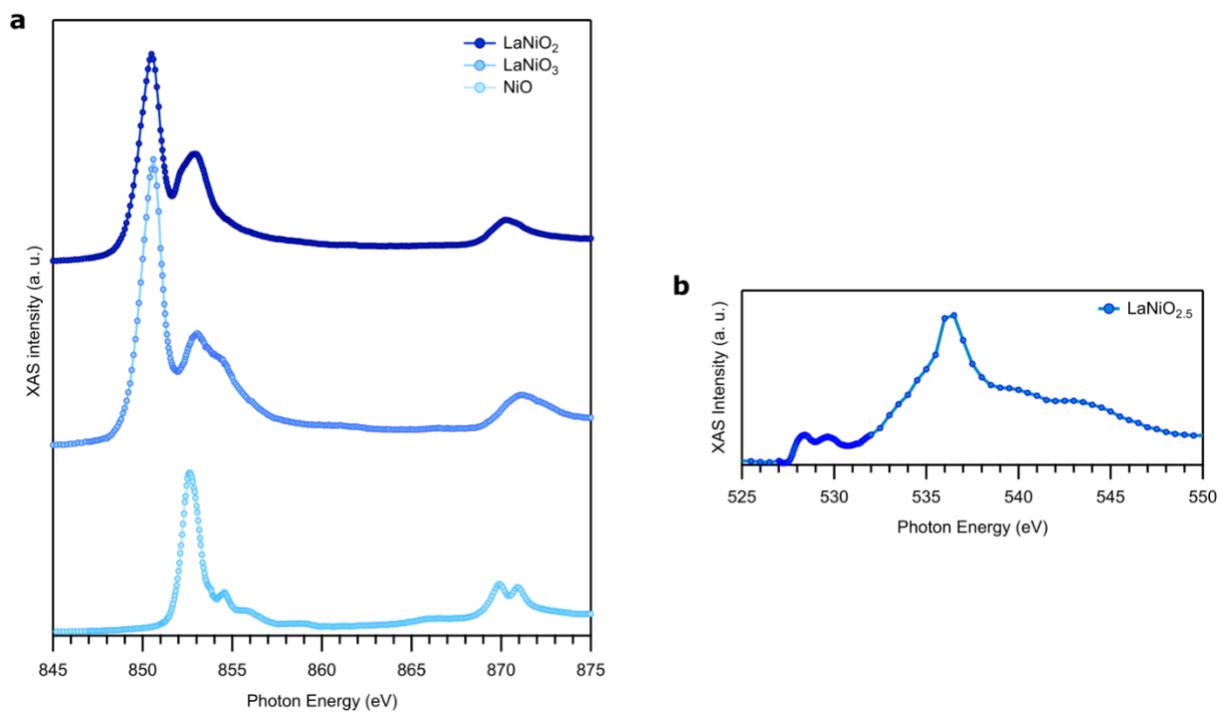

**Supplementary Figure 2 | Ni L-edge and O K-edge x-ray absorption spectra (XAS). A**, Normalized XAS of LaNiO3 and LaNiO2 across the Ni $L_{3,2}$-edge before subtraction of the La $M_4$ line. The XAS of NiO powder is also shown. Spectra are offset in vertical direction for clarity. **B**, Normalized XAS of LaNiO2.5 across the O K-edge.

| Unit eV | | Ni [0,0,0] | | | | | O[0, ½, 0] | | | O[½, 0, 0] | | | La [½,½,½] |
|---|---|---|---|---|---|---|---|---|---|---|---|---|---|
| | | $d_{z2}$ | $d_{x2-y2}$ | $d_{xy}$ | $d_{xz}$ | $d_{yz}$ | $p_z$ | $p_x$ | $p_y$ | $p_z$ | $p_x$ | $p_y$ | $d_{z2}$ |
| Ni [0,0,0] | $d_{z2}$ | -0.04 | | | | | | | -0.3 | | 0.3 | | -0.5 |
| | $d_{x2-y2}$ | | 0.70 | | | | | | -1.2 | | -1.2 | | |
| | $d_{xy}$ | | | -0.1 | | | | 0.7 | | | | 0.7 | |
| | $d_{xz}$ | | | | 0 | | | | | -0.8 | | | -0.1 |
| | $d_{yz}$ | | | | | 0 | 0.8 | | | | | | 0.1 |
| O [0, ½, 0] | $p_z$ | | | | | 0.8 | -2.32 | 0.2 | | | | | 0.4 |
| | $p_x$ | | | 0.7 | | | 0.2 | -2.34 | | | 0.3 | 0.2 | -0.2 |
| | $p_y$ | -0.3 | -1.2 | | | | | | -3.26 | | 0.6 | 0.3 | 0.4 |
| O [½, 0, 0] | $p_z$ | | | | -0.8 | | | | | -2.32 | | -0.2 | 0.4 |
| | $p_x$ | 0.3 | -1.2 | | | | | 0.3 | 0.6 | | -3.25 | | -0.4 |
| | $p_y$ | | | 0.7 | | | | 0.2 | 0.3 | -0.2 | | -2.35 | 0.2 |
| La [½,½,½] | $d_{z2}$ | -0.5 | | | -0.1 | 0.1 | 0.4 | -0.2 | 0.4 | 0.4 | -0.4 | 0.2 | 2.42 |

**Supplementary Table 1** Materials parameters for LaNiO$_2$ obtained from Wannier downfolding. The diagonal terms represent on-site energies. The off-diagonal terms represent the hopping between two orbitals. The shaded area only shows those parameters whose absolute value is larger than 0.1. All values are in units of eV. The triplet [i,j,k] appearing next to each orbital shows its relative position within a one Ni, LaNiO$_2$ tetragonal unit cell along the unit a, b, and c axes, respectively, with Ni at [0,0,0].

| Orbital configuration | | Approximate percentage | Dipole transition spin up (minority spin) | Dipole transition spin down (majority spin) |
|---|---|---|---|---|
| $d^9$ | $d^4\uparrow d^5\downarrow$ | 56% | Yes | No |
| $d^8R$ | $d^3R\uparrow d^5\downarrow$ | 24% | Yes | No |
| | $d^4\uparrow d^4R\downarrow$ | 14% | Yes | Yes |
| $d^7R^2$ | $d^3R\uparrow d^4R\downarrow$ | 6% | Yes | Yes |

**Supplementary Table 2:** Orbital configurations and their approximate percentage shown in the many-body ground state calculation. Here, R represents a rare earth electron in the La cage surrounding each Ni, in analogy to the standard ligand hole ($L$), due to hybridization between Ni and La. This is akin to a charge transfer, but from Ni to the rare earth, instead of oxygen to transition metal more commonly encountered in oxides.

| Hopping Parameters from Wannier Downfolding | | | |
|---|---|---|---|
| i | j | k | $t^{R}_{[i,j,k]}$ |
| 0 | 0 | 0 | 1.132 |
| 0 | 0 | 1 | -0.022 |
| 0 | 0 | 2 | -0.112 |
| 0 | 0 | 3 | 0.019 |
| 1 | 0 | 0 | -0.028 |
| 1 | 0 | 1 | -0.164 |
| 1 | 0 | 2 | 0.033 |
| 1 | 1 | 0 | -0.062 |
| 1 | 1 | 1 | 0.024 |
| 1 | 1 | 2 | 0.013 |
| 1 | 1 | 3 | -0.019 |
| 2 | 0 | 1 | 0.009 |
| 2 | 1 | 1 | 0.009 |
| i | j | k | $t^{Ni}_{[i,j,k]}$ |
| 0 | 0 | 0 | 0.267 |
| 1 | 0 | 0 | -0.355 |
| 1 | 1 | 0 | 0.090 |
| 2 | 0 | 0 | -0.043 |
| 0 | 0 | 1 | -0.043 |
| 1 | 1 | 1 | 0.013 |
| i | j | k | $t^{R-Ni}_{[i,j,k]}$ |
| 2 | 0 | 0 | -0.026 |
| 2 | 0 | 2 | 0.013 |

**Supplementary Table 3:** Tight-binding model parameters for the two-orbital model for LaNiO$_2$. The elements in the table show all the independent hopping parameters with a absolute magnitude larger than 0.008 eV. The triplet of integers [i,j,k] represents hopping between unit cells with a relative separation r = i a + j b + k c, where a, b, and c are unit vectors in the respective directions. Here, for simplicity we take a=b=c=1 and measure all momenta kx, ky, and kz, accordingly. Tetragonal symmetry dictates that for the rare earth- (*R*-) and Ni-band parameters, the hopping with triplet [i,j,k] would be equivalent to [-i,j,k], [i,-j,k], [i,j,-k], [-i,-j,k], [-i,j,-k], [i,-j,-k], [-i,-j,-k], and all combinations with i ↔ j, with a phase factor of 1 or -1 depending on the symmetry of the orbitals. The triplet [0,0,0] represents the orbital site energy ε$_0$. The La-centered and Ni-centered Wannier orbitals have relative unit cell coordinates [0.5, 0.5, 0.5] and [0.0, 0.0, 0.0], respectively,

which only affects the equivalent triplets for the cross-orbital-hybridization terms $t^{R-Ni}$. The fermi energy $E_F$ is at 0 eV.

**Supplementary References**